\documentclass[12pt]{article}
\usepackage{sc3conf}
\usepackage{amsfonts}

\begin{document}
\raggedbottom

\title{Some recent results in calculating the Casimir energy at
zero and finite temperature}

\authors{V.V.~Nesterenko}
%  T.T.~Them,\adref{2,3}
% and O.O.~Others\adref{1,3}}

\addresses{Bogoliubov Laboratory of Theoretical Physics,
 Joint Institute for Nuclear Research, \\ Dubna, 141980, Russia.}
%  \nextaddress \2ad Their address (my old one)
%  \nextaddress \3ad Another address (maybe my future one).}

\maketitle

\begin{abstract}
The survey summarizes briefly the results obtained recently in the
Casimir effect studies considering the following subjects:
   i) account of the material characteristics of the media and
their influence on the vacuum energy (for example, dilute
dielectric ball);
   ii) application of the spectral geometry methods for investigating
the vacuum energy of quantized fields with the goal to gain some
insight, specifically, in the geometrical origin of the
divergences that enter the vacuum energy and to develop the
relevant renormalization procedure;
  iii) universal method for calculating the high temperature dependence
of the Casimir energy  in terms of heat kernel coefficients.
 \end{abstract}
\section{Introduction}
The progress in  calculation of the Casimir energy is rather slow.
In 1948 Casimir derived the vacuum electromagnetic energy for the
most simple boundary conditions, i.e., for two parallel perfectly
conducting plates placed in vacuum (the relevant references can be
found in an excellent book~\cite{book}). Dielectric properties of
the media separated by plane boundaries did not add new
mathematical difficulties. However, the first result on the
calculation of the Casimir energy for nonflat boundaries was
obtained only in 1968. By computer calculations, which then lasted
three  years, Boyer found the Casimir energy of a perfectly
conducting spherical shell. Account of dielectric and magnetic
properties of the media in calculations of the vacuum energy for
nonflat interface leads to new principal difficulties or, more
precisely, to a new structure of divergencies. Nonsmoothness of
the boundary (corners, edges and so on) also contributes to the
vacuum divergences. The temperature behavior of the Casimir effect
is a problem of independent interest.

 The review  summarizes briefly the following results obtained
recently in the Casimir calculations: dilute dielectric ball
(Sec.~2), boundary with corners (Sec.~3), and high temperature
expansion of the Casimir energy in terms of  heat kernel
coefficients (Sec.~4).

\section{Casimir energy of a dilute dielectric ball}
Calculation of the Casimir energy of a dielectric ball has a
rather long history starting 20 years ago~\cite{1}. However only
recently the final result was obtained for a dilute dielectric
ball at zero \cite{2,LSN} and finite \cite{NLS,Barton}
temperature. Here we summarize briefly the derivation of the
Casimir energy of a dilute dielectric ball by making use of the
mode summation method and the addition theorem for the Bessel
functions instead of the uniform asymptotic expansion for these
functions \cite{LSN,NLS}.

%\section{The Basic Steps of Calculations and the Results}
\noindent A  solid ball of radius $a$ placed in an unbounded
uniform medium is considered. The contour integration technique
\cite{LSN}  gives ultimately the following representation for the
Casimir energy of the ball
\begin{equation}
 \label{1}
  E=-\frac{1}{2\pi a}\sum_{l=1}^{\infty}(2l+1)\int_0^{y_0}dy\,
 y\,\frac{d}{dy}\ln\left[W_l^2(n_1y, n_2y)-\frac{\Delta n^2}{4}\,
 P_l^2(n_1y, n_2y)\right]\,,
\end{equation}
where
\begin{eqnarray}
  W_l(n_1y,
n_2y)&=&s_l(n_1y)e_l^{\prime}(n_2y)-s_l^{\prime}(n_1y)e_l(n_2y)\,,
  \nonumber \\
  P_l(n_1y,
n_2y)&=&s_l(n_1y)e_l^{\prime}(n_2y)+s_l^{\prime}(n_1y)e_l(n_2y)\,,
\nonumber
\end{eqnarray}
and $s_l(x)$, $e_l(x)$ are the modified Riccati-Bessel functions,
$n_1, n_2 $ are the refractive indices of the ball and of its
surroundings, $\Delta n= n_1-n_2$.

Analysis of divergences \cite{LSN} leads to the following
algorithm for calculating the vacuum energy (\ref{1}) in the
$\Delta n^2$-approximation. First, the $\Delta n^2$--contribution
should be found, which is given by the sum $\sum_lW_l^2$. Upon
changing its sign to the opposite one, we obtain the contribution
generated by $W_l^2$, when this function is in the argument of the
logarithm. The $P^2_l$-contribution into the vacuum energy is
taken into account by expansion of Eq. (\ref{1}) in terms of
$\Delta n^2$.

 Applying the addition theorem for the Bessel
functions~\cite{AS}
\[
  \sum_{l=0}^{\infty}(2l+1)[s_l^{\prime}(\lambda
  r)e_l(\lambda\rho)]^2=\frac{1}{2r \rho}\int_{r-\rho}^{r+\rho}
  \left(\frac{1}{\lambda}\,\frac{\partial{\cal D}}{\partial r}
 \right)^2R\;dR
\]
with
\[
{\cal D} =\frac{\lambda r\rho}{R}\, e^{-\lambda R}, \quad
  R=\sqrt{r^2+\rho^2-2r\rho\cos\theta}
\]
one arrives at the result
\[
  E=\frac{23}{384}\frac{\Delta n^2}{\pi a}
  =\frac{23}{1536}\,\frac{(\varepsilon_1-
  \varepsilon_2)^2}{\pi a}{,} \quad  \varepsilon_i=n^2_i, \quad i=1,2\,{.}
\]

Extension to finite temperature $T$ is accomplished by
substituting the $y$-integra\-tion in (\ref{1}) by summation over
the Matsubara frequencies $ \omega_n=2\pi nT$. When considering
the  low temperature  behavior of the thermodynamic functions of a
dielectric ball   the term proportional to $T^3$ in our paper
\cite{NLS} was lost. It was due to the following. We have
introduced the summation over the Matsubara frequencies in Eq.\
(3.20) under the sign of the $R$-integral. Here we show how to do
this summation in a correct way.

In the $\Delta^2$-approximation the last term in Eq.\ (3.20) from
the article \cite{NLS}
\begin{equation}
\label{eq-2} \overline {U}_W(T)=2 T\Delta n^2 \sum_{n=0}^\infty
\!{}^{'}w^2_n\int_{\Delta n}^2\frac{e^{-2w_nR}}{R}\,dR{,} \quad
w_n=2 \pi na T
\end{equation}
can be represented in the following form
\begin{equation}
\label{eq-3} \overline {U}_W(T)=-2 T\Delta n^2
\sum_{n=0}^\infty\!{}^{'} w^2_n \,E_1(4w_n){,}
\end{equation}
where $E_1(x)$ is the exponential-integral function \cite{AS}. Now
we accomplish the summation over the Matsubara frequencies by
making use of the Abel-Plana formula
\begin{equation}
\label{eq-4} \sum_{n=0}^\infty\!{}^{'} f(n) =\int_0^\infty
f(x)\,dx+i\int_0^\infty \frac{f(ix)-f(-ix)}{e^{2\pi x}-1}\;dx{.}
\end{equation}
The first term in the right-hand side of this equation gives the
contribution independent of the temperature, and the net
temperature dependence is produced by the second term in this
formula. Being interested in the low temperature behavior of the
internal energy we substitute into the second term in Eq.\
(\ref{eq-4}) the following  expansion of the function $E_1(z)$
\begin{equation}
\label{eq-5} E_1(z) =-\gamma -\ln z- \sum^\infty_{k=1}\frac{(-1)^k
z^k}{k\cdot k!},\quad |\arg z |<\pi {,}
\end{equation}
where $\gamma $ is the Euler constant \cite{AS}. The contribution
proportional to $T^3$ is produced by the logarithmic term in the
expansion (\ref{eq-5}). The higher powers of $T$ are generated by
the respective terms in the sum over $k$ in this formula $(t=2\pi
a T)$
\begin{equation}
\label{eq-6} \overline {U}_W(T)=\frac{\Delta n^2}{\pi a}\left (
-\frac{1}{96}+ \frac{\zeta (3)}{4\pi ^2} t^3 -\frac{1}{30}t^4
+\frac{8}{567} t^6 -\frac{8}{1125}t^8+{\cal O}(t^{10})\right ) {.}
\end{equation}
All these terms, safe for $2 \zeta (3) \Delta n^2 a^2T^3$, are
also reproduced by the last term in Eq.\ (3.31) in our paper
\cite{NLS} (unfortunately additional factor 4 was missed there)
\[
\frac{\Delta n^2}{8}T\cdot 4\, t^2\int ^2_{\Delta n}\frac {dR}{R}
\frac{\coth (tR)}{\sinh^2 (tR)}{.}
\]
Taking all this into account we arrive at the following low
temperature behavior of the internal Casimir energy of a dilute
dielectric  ball
\begin{equation}
\label{eq-7} U(T)= \frac{\Delta n^2}{\pi a}\left ( \frac{23}{384}
+\frac{\zeta(3)}{4\pi^2}t^3 -\frac{7}{360}t^4 +\frac{22}{2835}t^6
-\frac{46}{7875}t^8 +{\cal O}(t^{10})
 \right ){.}
\end{equation}
The relevant  thermodynamic relations give the following low
temperature expansions for free energy
\begin{equation}
\label{eq-8} F(T)=\frac{\Delta n^2}{\pi a}\left (
\frac{23}{384}-\frac{\zeta (3)}{8\pi ^2}t^3+\frac{7}{1080}t^4
 -\frac{22}{14175}t^6+\frac{46}{55125}t^8+{\cal O}(t^{10})
\right )
\end{equation}
and for entropy
\begin{equation}
\label{eq-9} S(T)=-\frac{\partial F}{\partial T}=\Delta n^2 \left
( \frac{3\zeta (3)}{4\pi ^2}t^2-\frac{7}{135}t^3
+\frac{88}{4725}t^5- \frac{736}{55125}t^7+ {\cal O}(t^9) \right
){.}
\end{equation}

The range of applicability of the  expansions (\ref{eq-7}),
(\ref{eq-8}), and (\ref{eq-9}) can be roughly estimated in the
following way. The curve $S(T)$ defined by Eq.\ (\ref{eq-9})
monotonically goes up when the dimensionless temperature $t =2\pi
a T$ changes from 0 to $t \sim 1.0$. After that  this curve
sharply goes down to the negative values of $S$. It implies  that
Eqs.\ (\ref{eq-7}) -- (\ref{eq-9}) can be used in the region
$0\leq t < 1.0$. The $T^3$-term in Eqs.\ (\ref{eq-7}) and
(\ref{eq-8}) proves to be principal because it gives the first
positive term in the low temperature expansion for the  entropy
(\ref{eq-9}). It is worth noting, that the exactly the same
$T^3$-term, but with opposite sign, arises in the high temperature
asymptotics of free energy in the problem at hand (see Eq.\ (4.30)
in Ref.~\cite{BNP}).

For large temperature $T$ we found~\cite{NLS}
\begin{equation}
\label{eq-10}
 U(T) \simeq  \frac{\Delta n^2}{8}\, T {,}\;\;
 F(T)  \simeq  -\frac{\Delta n^2}{8}\, T\left [\ln (aT)-c\right ]{,}\;\;
 S(T)\simeq \frac{\Delta n^2}{8}\left [
\ln (aT)+c+1 \right ] {,}
\end{equation}
where $c$ is a constant \cite{Barton,BNP} $ c=\ln 4 +\gamma
-{7}/{8}\,{.}$ Analysis of Eqs.\ (3.20) and (3.31) from the paper
\cite{Barton} shows that there are only exponentially suppressed
corrections to the leading terms (\ref{eq-10}).

Summarizing we conclude that now there is a complete agreement
between the results of  calculation of  the Casimir thermodynamic
functions for a dilute dielectric ball carried out in the
framework of two different approaches:  by the mode summation
method \cite{LSN,NLS} and by perturbation theory for quantized
electromagnetic field, when  dielectric ball is considered as a
perturbation in unbounded continuous surroundings \cite{Barton}.

%  Some text, and possibly some equations, e.g.,
%  \begin{equation}
%  \label{eq:1}
%  x^n + y^n = z^n,\qquad n\geq 3,\quad x,y,z\in\mathbb{N}.
% \end{equation}
% Another crucial equation is given in what follows (see
% Eq.~(\ref{eq:2}) in Sec.~\ref{sec:333}).

\section{Spectral geometry and vacuum energy}
\label{sec:333}
 In spite of a quite long history of the Casimir
effect (more than 50 years) deep understanding and physical
intuition in this field are still lacking. The main problem here
is the separation of net finite effect from the divergences
inevitably present in the Casimir calculations. A convenient
analysis of these divergences gives the heat kernel technique,
namely, the coefficients of the asymptotic expansion of the heat
kernel.

Keeping in mind the elucidation of the origin of these divergences
in paper \cite{s-c} the vacuum energy of electromagnetic field has
been calculated for a semi-circular infinite cylindrical shell.
This shell is obtained by crossing an infinite cylinder by a plane
passing through  its symmetry axes. In the theory of waveguides it
is well known that a semi-circular waveguide has the same
eigenfrequencies as the cylindrical one but without degeneracy
(without doubling) and safe for one frequency series.
Notwithstanding the very close spectra, the vacuum divergences in
these problems prove to be drastically different, so the zeta
function technique does not give a finite result for a
semi-circular cylinder unlike for a circular one.

 It was revealed that the origin of these divergences is
the corners in the boundary of semi-circular cylinder~\cite{NPD}.
In terms of the heat kernel coefficients, it implies that the
coefficient $a_2$ for a semi-circular cylinder does not vanish due
to these corners.

However in the 2-dimensional (plane) version of these problems the
origin of nonvanishing $a_2$ coefficient for a semicircle  is the
contribution due to the curvature of the boundary, while the
corner contributions  to $a_2$ in 2 dimensions are cancelled.

Different geometrical origins of the vacuum divergences in the
two- and three-dimensional versions of the boundary value problem
in question evidently imply  the impossibility of obtaining a
finite and unique value of the Casimir energy  by taking advantage
of the atomic structure of the boundary or its quantum
fluctuations. It is clear, because any physical reason of the
divergences should hold simultaneously in the two- and
three-dimensional versions of a given boundary configuration.

\section{High temperature asymptotics
of vacuum energy in terms of  heat kernel coefficients}
\label{sec:444} The Casimir calculations at finite temperature
prove to be a nontrivial problem specifically for boundary
conditions with nonzero curvature.  For this goal a powerful
method of the zeta function technique and the heat kernel
expansion can be used. For obtaining the high temperature
asymptotics of the thermodynamic characteristics it is sufficient
to know the heat kernel coefficients and the determinant   for the
spatial part of the operator governing the field dynamics.
 This is an essential merit
of this approach~\cite{BNP}.

Starting point is the general high temperature expansion of the
free energy in terms of the heat kernel coefficients
$a_n$~\cite{DK}
\begin{eqnarray}
F(T)&\simeq&-\frac{T}{2}\zeta'(0)-a_0\frac{T^4}{\hbar^3}\,
\frac{\pi^2}{90}-\frac{a_{1/2}T^3}{4\pi^{3/2} \hbar^2 } \zeta_R(3)
-\frac{a_1}{24}\frac{T^2}{\hbar}
+\frac{a_{3/2}}{(4\pi)^{3/2}}T\ln\frac{\hbar}{T} \nonumber \\
&& -\frac{a_2}{16\pi^2}\hbar \left[\ln\left(\frac{\hbar}{4 \pi
T}\right)+\gamma\right]-\frac{a_{5/2}}{(4\pi)^{3/2}}\frac{\hbar^2}{24
T} \nonumber\\&&-T\sum_{n\geq
3}\frac{a_n}{(4\pi)^{3/2}}\left(\frac{\hbar}{2 \pi
T}\right)^{2n-3} \,
 \Gamma(n-3/2)\,\zeta_R(2 n-3)\,{.}
 \label{F}
\end{eqnarray}
Here $\gamma$ is the Euler constant and $\zeta_{R}(s)$ is the
Riemann zeta function. The quantities under the logarithm sign in
expansion (\ref{F}) are dimensional, but upon collecting similar
terms with account for the logarithmic ones in $\zeta'(0)$ it is
easy to see that  finally the  logarithm function has a
dimensionless argument.

The first term in the asymptotics of the free energy in Eq.\
(\ref{F}) is referred to as a pure entropic contribution. Its
physical origin is till now not elucidated. The entropic term is a
pure classical quantity because it does not involve the Planck
constant $\hbar$. This classical contribution to the asymptotics
seems
 to be derivable
without appealing to the notion of quantized electromagnetic
field.

The heat kernel coefficients needed for construction of the
expansion (\ref{F}) will been  calculated as the residua of the
corresponding zeta functions. For the boundary conditions under
consideration the explicit expressions for the zeta functions have
been derived in~\cite{LNB}.

{\it A perfectly conducting spherical shell of radius $R$ in
vacuum.} The first six heat kernel coefficients in this problem
are:
\begin{eqnarray}
a_0=0, \quad a_{1/2}=0,\quad a_1=0, \quad
\frac{a_{3/2}}{(4\pi)^{3/2}}= \frac{1}{4}, \quad a_2=0,\quad
\frac{a_{5/2}}{(4\pi)^{3/2}}=\frac{c^2}{160\,R^2}. \label{eq_2}
\end{eqnarray}
Furthermore
\begin{equation}
a_j=0,\qquad j=3,4,5,  \dots\, {.} \label{eq3}
\end{equation}
 The exact value of  $\zeta' (0)$ is
derived in~\cite{BNP}
\begin{equation}
\zeta' (0)= 0.38429+\frac{1}{2}\ln\frac{R}{c}. \label{eq4}
\end{equation}
As a result we have the following high temperature asymptotics of
the  free energy
\begin{equation}
F (T)=-\frac{T}{4}\left({\displaystyle 0.76858}+\ln \tau+
 \frac{1}{960\tau ^2}\right )+{\cal O }(T^{-3}),
\label{eq5}
\end{equation}
where $\tau =RT/(\hbar c)$ is the dimensionless `temperature'. The
expression (\ref{eq5}) exactly reproduces the  asymptotics
obtained in \cite{BD} by making use of the multiple scattering
technique (see Eq.~(8.39) in that paper).

{\it A compact ball with $c_1=c_2$.} In this case the spherical
surface  delimits the media with ``relativistic invariant''
characteristics i.e., the velocity of light is the same inside and
outside the ball. Here there naturally  arises \cite{BrNP} a
dimensionless parameter $ \xi^2=\left(\frac{\varepsilon_1-
\varepsilon_2}{\varepsilon_1+\varepsilon_2}\right)^2=
\left(\frac{\mu_1-\mu_2}{\mu_1+\mu_2}\right)^2, $ where
$\varepsilon_1$ and $\varepsilon_2$ ($\mu_1$ and $\mu_2$) are
permittivities (permeabilities) inside and outside the ball. As
usual we perform the calculation in the first order of the
expansion with respect to $\xi^2$.

The zeta function for this boundaries, obtained in  \cite{LNB},
affords the exact values of heat kernel coefficients up to $a_3$
\begin{equation}
a_0= a_{1/2}= a_1=0,\quad a_{3/2}=2\pi^{3/2}\xi^2,  \quad
 a_2=0, \quad  a_{5/2}=0, \quad
a_3=0.\label{eq4_14}
\end{equation}
The zeta determinant in this problem turns out to be given  by
multiplication of the content of the second parentheses in Eq.\
(\ref{eq4})  by $\xi^2$
\begin{equation}
\zeta'(0) =\xi^2\left(0.35676+\frac{1}{2}\ln\frac{R}{c}\right).
\label{eqbz}
\end{equation}
The high temperature asymptotics for free energy reads

\begin{equation}
F(T)= -\xi^2\frac{T}{4}\left (0.71352+ \ln \tau \right )+{\cal
O}(T^{-3}){.} \label{eq4_17}
\end{equation}
The asymptotics (\ref{eq4_17})  completely coincide with analogous
formula obtained in \cite{NLS} by the mode summation method
combined with the addition theorem for the Bessel functions.

{\it A perfectly conducting cylindrical shell.} The heat kernel
coefficients are
\begin{equation}
a_0= a_{1/2}= a_1= a_2=0,\quad
\frac{a_{3/2}}{(4\pi)^{3/2}}=\frac{3}{64\,R}, \quad
\frac{a_{5/2}}{(4\pi)^{3/2}}=\frac{153}{8192}\frac{c^2}{R^3}.
\end{equation}
The zeta function determinant in this problem is calculated
in~\cite{BNP}
\begin{equation}
\zeta'(0)=\frac{0.45711}{R}+\frac{3}{32\,R}\,\ln\frac{R}{2\,c}.
\label{eq5_6}
\end{equation}
The free energy behavior at high temperature is the following
\begin{equation}
F(T)=-\frac{T}{R} \left (0.22856 +\frac{3}{64} \ln\frac{\tau
}{2}-\frac{51}{65536\tau ^2}\right ) +{\cal O}(T^{-3}).
\label{fcs}
\end{equation}
 The high temperature asymptotics  of the electromagnetic
free energy in presence of perfectly conducting cylindrical shell
was investigated  in  \cite{BD}. To make the comparison handy let
us rewrite  their result as follows
\begin{equation}
F(T)\simeq-\frac{T}{R}\left (0.10362+\frac{3}{64R}
\ln\frac{\tau}{2}\right ). \label{15a}
\end{equation}
The discrepancy between the  terms linear  in $T$ in Eqs.\
(\ref{fcs}) and (\ref{15a}) is due to the double scattering
approximation used in \cite{BD} (see also below). Our approach
provides  an opportunity to calculate the exact value of this term
(see Eq.\  (\ref{fcs})).

Thus we have  demonstrated efficiency and universality of the high
temperature expansions in terms of the heat kernel coefficients
for the Casimir problems with spherical and cylindrical
symmetries. All the known results in this field are reproduced  in
a uniform approach and in addition  a few new asymptotics are
derived~\cite{BNP}.

\section{Conclusions}
The inferences concerning the individual subjects of this brief
review have been done in respective sections. Here we would like
only to note, that in order to cast the theory of the Casimir
effect  to a complete form further studies are certainly needed.

 {\bf Acknowledgements.} The
results presented above have been obtained in collaboration with
I.G.~Pirozhenko, G.~Lambiase, G.~Scarpetta, M.~Bordag, and
J.~Dittrich. The author is grateful to  them. The work was
supported in part by  RFBR (Grant No.\ 00-01-00300) and by  ISTC
(Project No.\ 840).

\end{document}